\documentclass{aa}      
\usepackage{graphicx}
\usepackage{txfonts}
\usepackage{natbib}
\usepackage[bookmarks=false, colorlinks=true, citecolor=blue, linkcolor=blue]{hyperref}
\def\kms{\,km\,s$^{-1}$}

\begin{document} 


\title{Observations of 12.2\,GHz methanol masers towards northern high-mass protostellar objects} 

   \author{M. Durjasz
          \inst{} \href{https://orcid.org/0000-0001-7952-0305}{\includegraphics[scale=0.5]{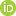}}
          \and
          M. Szymczak
          \inst{} \href{https://orcid.org/0000-0002-1482-8189}{\includegraphics[scale=0.5]{orcid.png}}
          \and
          P. Wolak
          \inst{} \href{https://orcid.org/0000-0002-5413-2573}{\includegraphics[scale=0.5]{orcid.png}}
          \and
          A. Bartkiewicz
          \inst{} \href{https://orcid.org/0000-0002-6466-117X}{\includegraphics[scale=0.5]{orcid.png}}
          }

   \institute{Institute of Astronomy, Faculty of Physics, Astronomy and Informatics, Nicolaus Copernicus University, Grudziadzka 5, 87-100 Torun, Poland}

  \date{Received XXX / Accepted XXX}

 
  \abstract
{Class II methanol masers at 6.7 and 12.2\,GHz occur close to high-mass young stellar objects (HMYSOs). When they are observed simultaneously, such studies may contribute to refining the characterisation of local physical conditions.}
{We aim to search for the 12.2\,GHz methanol emission in 6.7\,GHz methanol masers that might have gone undetected in previous surveys of northern sky HMYSOs, mainly due to their variability. Contemporaneous observations of both transitions are used to refine the flux density ratio and examine the physical parameters.} 
{We observed a sample of 153 sites of 6.7\,GHz methanol maser emission  in the 12.2\,GHz methanol line with the Torun 32\,m radio telescope, using the newly built X-band receiver.}
{The 12.2\,GHz methanol maser emission was detected in 36 HMYSOs, with 4 of them detected for the first time. The 6.7\,GHz to 12.2\,GHz flux density ratio for spectral features of the contemporaneously observed sources has a median value of 5.1, which is in agreement with earlier reports. The ratio differs significantly among the sources and for the periodic source G107.298+5.639 specifically, the ratio is weakly recurrent from cycle to cycle, but it generally reaches a minimum around the flare peak. This is consistent with the stochastic maser process, where small variations in the physical parameters along the maser path can significantly affect the ratio.
A comparison of our data with historical results (from about ten years ago) implies significant ($>50$\%) variability for about 47\% and 14\% at 12.2\,GHz and 6.7\,GHz, respectively. This difference can be explained via the standard model of methanol masers.
}
{}


\titlerunning{12.2\,GHz methanol maser survey towards 6.7\,GHz sources}   
\authorrunning{M. Durjasz et al.}

\maketitle

\section{Introduction}
Masers play a significant role in the study of the interstellar medium. The methanol molecule has attracted much attention, from its first detection half a century ago \citep{barrett1971} to the present, given the multitude of observed transitions. One of the breakthroughs in the field came with the detection of the strongest, and most pervasive, lines at 12.2\,GHz \citep{batrla1987} and 6.7\,GHz \citep{menten1991}, which provided powerful tools for identifying high-mass young stellar objects (HMYSOs) and exploring the physical conditions and gas kinematics of their surroundings. A number of new methanol transitions in the centimetre \citep{breen2019,macleod2019} and millimetre \citep{brogan2019} ranges have been discovered very recently in objects that have undergone major accretion events.

The 6.7 and 12.2\,GHz lines are referred to as class II methanol maser transitions \citep{batrla1987,menten1991} that are characterised by radiative pumping \citep{cragg2005} and a close association with sources of strong radiation. They are usually found in the vicinity of HYMSOs, as revealed by several surveys at 6.7\,GHz (e.g. \citealt{caswell1995a, ellingsen1996, pandian2007, szymczak2012, green2010, breen2015}). Several searches for the 12.2\,GHz methanol maser emission were carried out prior to the discovery of the 6.7 GHz line, mainly towards OH maser sites \citep{norris1987,koo1988,kemball1988,caswell1993,macleod1993}. Thereafter, all the 12.2\,GHz surveys were done towards the 6.7\,GHz sources \citep{gaylard1994,caswell1995b,blaszkiewicz2004,breen2010,breen2012b,breen2012a,breen2014,breen2015}, as the 12.2\,GHz line excitation conditions closely follow those at 6.7\,GHz (\citealt{cragg2002}). Almost all of these surveys did not cover the entire northern sky, thus, in this study, we  attempt to supplement the observations with objects of the northern hemisphere. 

No conclusive data is available on the variability patterns of the 12.2\,GHz methanol masers. Little or no variability was noticed on a timescale of seven months in eight bright ($>$100\,Jy) 12.2\,GHz maser sources \citep{mccutcheon1988}, whereas in the other five bright sources, the emission varied internally by less than 15-20\% over a four-year period with the exception of single weak features in two sources \citep{macleod1993}. In turn, \citet{caswell1993} reported the 12.2\,GHz flux density variations on a timescale of weeks and later found that at least a quarter of the features exhibit intensity variations larger than 10\% on a timescale of 4 yr \citep{caswell1995b}
suggesting that most quiescent 12.2\,GHz masers are saturated. In this paper, we present the results of multi-epoch 12.2\,GHz observations, along with a comparison to historical observations, for 153 sources. We include several contemporaneously monitored 6.7\,GHz methanol maser counterparts.

\section{Observations} \label{sec:obser}
Observations were carried out in the period from 2019 August to 2020 February using the Torun 32\,m radio telescope. The telescope has a half-power beam width of about 3\arcmin\, at 12.2\,GHz and rms pointing errors of about 10\arcsec. The adopted rest frequency was 12.178597\,GHz \citep{muller2004}. We used the newly built X-band receiver  \citep{pazderski2018}, which is a dual-polarisation cooled receiver with amplifiers of typical noise temperature of 5\,K.  Here, the system temperature was about 30\,K. The  antenna  gain  was estimated to be about 0.06\,K\,Jy$^{-1}$ from observations
of calibrators DR21 and 3C123 adopting flux densities of
20.0 and 6.3\,Jy, respectively \citep{ott1994}. No gain elevation correction was applied; this can contribute less than 4\% to the error budget of the flux density. The uncertainty in the flux density scale is about 15\%. 

The observations were made in frequency-switching mode. An autocorrelation spectrometer was configured to record 4096 channels either with 4\,MHz or 8\,MHz bandwidth for each circularly polarised signal yielding the channel spacing of 0.024 or 0.048\kms, respectively. The velocity extent of each observation is $\pm$45 or $\pm$90\kms\, with respect to the local standard of rest. 
A typical 3$\sigma$ noise level in the final spectra with the higher spectral resolution was 1.5 Jy. The stability of the system was regularly checked with observations of G188.946+0.886 that
show no variability to a limit of 15\% during our observing interval, while its 12.2\,GHz spectrum shape remained almost unchanged after $\sog$10\,yr when observed by \citet{breen2012b}.
Observations of G188.946+0.886 during the 2000-2008 revealed periodic  (395$\pm$8\,d) variability with relative amplitudes of 0.5 and 0.7 at 6.7 and 12.2\,GHz, respectively, but periodic changes at 12.2\,GHz were barely visible prior to 2005 (\citealt{goedhart2014}, their Figs.\,9 and 10). In the period from 2009 to 2013, the relative amplitude of periodic variations at 6.7\,GHz decreased to about 0.2 \citep{szymczak2018}, whereas our unpublished 6.7\,GHz observations, carried out contemporaneously with the present 12.2\,GHz survey, show a variability that is lower than 10\% We argue that the variability of this object at 12\,GHz does not exceed the uncertainty of our measurements.
For selected sources, contemporaneous ($\le\pm$7\,d) observations were carried out in the 6.7\,GHz methanol transition following the procedure described in \citet{szymczak2018}.

We selected bright 6.7\,GHz sources \citep{breen2015, szymczak2012, hu2016, green2010} that are observable from the northern hemisphere and mostly with a declination of $>0\degr$
to avoid a degradation of the telescope sensitivity due to radio frequency interference from commercial satellites. The final sample consists of 153 sources.

\section{Results}
\begin{figure}
\includegraphics[scale=0.65]{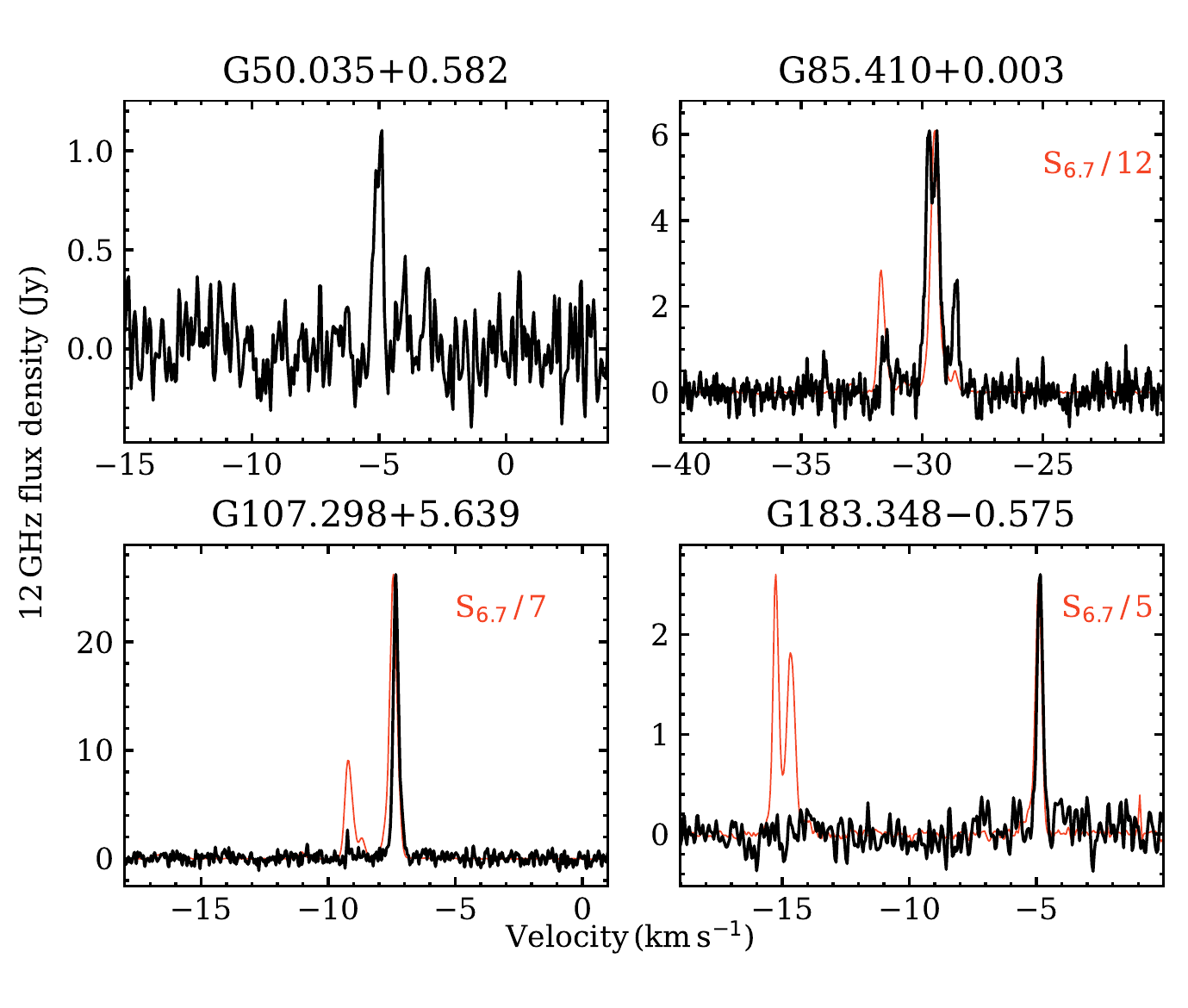}
\caption{Spectra of the newly detected 12.2\,GHz methanol maser sources. Spectra of 6.7\,GHz methanol masers (red) are also shown when taken on the same day, with exception of G183.348$-$0.575, where the interval between observations was 41\,d. For comparison purposes,
the scale of 6.7\,GHz flux density was reduced by the factor given in the upper right corner.
\label{fig:spectras1}}
\end{figure}

The 12.2\,GHz maser emission was detected in 36 objects (Table\,\ref{tab:dec_list}), including four new sources. Figures\,\ref{fig:spectras1} and \ref{fig:specrtas_nonewdec} present the spectra of newly detected and previously known sources, respectively. The list of non-detections is also given in Table\,\ref{tab:non-det}.

\subsection{Newly detected 12.2\,GHz sources}
\textit{G50.035+0.582}. This source has a single 12.2\,GHz feature at $-$5.1\kms\,, with S$_{12.2}$ = 1.1\,Jy, which coincides with the strongest feature of the 6.7\,GHz methanol maser \citep{szymczak2012,breen2015}. There was no 12.2\,GHz maser emission detected in 2010 March with the 5$\sigma$ sensitivity of 0.84\,Jy \citep{breen2016}. This implies a variability of $\ge$30\% on a timescale of 9.5\,yr. 

\textit{G85.410+0.003}. The 12.2\,GHz spectrum is comprised of three features at $-$31.5, $-$29.5, and $-$28.6\kms. These are coincident with their 6.7\,GHz maser counterparts, within 0.1\kms, observed in the same epoch. 
In instances where a single bright feature at $V_{12.2}$ = $-$29.5\kms\ was originally detected, we now report two features separated by 0.3\kms.
A comparison with the Effelsberg 100\,m spectrum taken on May 19, 2020  (Yen and Menten, private communication at M2O\footnote{www.masermonitoring.com}) implies significant changes in the profile shape and flux density of individual features by a factor of 2$-$3 over four months.

\textit{G107.298+5.639}. The single velocity feature is matched in velocity to the strongest 6.7\,GHz feature at $-$7.35\kms\, \citep{olech2020} during all three observed cycles, which are discussed in detail in Sect.\,\ref{sec:g107_variability}.  Hereafter, G107.298+5.639 is referred to as G107.

\textit{G183.348$-$0.575}. The 12.2\,GHz maser emission consists of a single feature at $-$4.9\kms\, with a peak flux density of 2.5\,Jy, which  coincides in velocity with the 6.7\,GHz maser feature. There was no 12.2\,GHz maser emission seen for the blueshifted component of the 6.7\,GHz emission near $-$15\kms. 

\setlength{\tabcolsep}{4pt}
\begin{table*}
\centering
\caption{List of detected 12.2\,GHz methanol maser sources. Column 1 presents the methanol maser galactic coordinates, columns 2-3 give equatorial coordinates of the 6.7\,GHz methanol masers \citep{breen2016, szymczak2012, hu2016}. Columns 4-8 present characteristics of the obtained spectra; the velocity of the strongest peak relative to the local standard of rest ($V_{\mathrm{peak}}$), peak flux density ($S_{\mathrm{peak}}$), velocity range of the maser emission ($\Delta V$), integrated flux density ($S_{\mathrm{int}}$), and epoch of observation, respectively. For G107.298+5.639 the epoch of the strongest flare was selected. The last column list the references to previous detections. 
\label{tab:dec_list}}
\begin{tabular}[c]{lcccccccc}
\hline
Name (l   b)  &  $\alpha$ (J2000) &  $\delta$ (J2000) & $V_{\mathrm{peak}}$ & $S_{\mathrm{peak}}$  & $\Delta V$ & $S_{\mathrm{int}}$ & Epoch & References\\
(\degr \hspace{0.5cm} \degr) & (h \hspace{0.2cm}  m \hspace{0.2cm}   s) & (\degr \hspace{0.2cm} \arcmin  \hspace{0.2cm} \arcsec) & (km\,s$^{-1}$) & (Jy) & (km\,s$^{-1}$) & (Jy\,km\,s$^{-1}$) & (MJD)\\
\hline
 G30.225$-$0.180  & 18 47 08.30 & $-$02 29 28.90 &  113.35 & 6.7 & 107;116 & 8.76 & 58723 & 9,11\\
 G32.744$-$0.075  & 18 51 21.87 & $-$00 12 05.00 &   30.55 & 3.7 & 29;40 & 5.81 & 58769 & 6,9,11,14\\
 G33.641$-$0.228  & 18 53 32.56 &    00 31 39.18 &   60.26 & 37.8 & 58;62 & 13.13 & 58868 & 11,14  \\
 G35.132$-$0.744  & 18 58 06.14 &    01 37 07.50 &   29.72 & 11.8 & 27;32 & 10.83 & 58872 &  14 \\
 G35.197$-$0.743  & 18 58 13.05 &    01 40 35.70 &   28.26 & 39.1 & 27;32 & 31.7 & 58873 & 6,7,9,11,14 \\
 G35.200$-$1.736  & 19 01 45.54 &    01 13 32.60 &   45.16 & 11.9 & 43;47 & 4.23 & 58873 & 2,4,5,6,9,10,11,14 \\
 G36.115+0.552    & 18 55 16.79 &    03 05 05.41 &   75.08 & 2.4 & 73;78 & 1.32 & 58720 & 14 \\
 G37.043$-$0.035  & 18 59 04.41 &    03 38 32.80 &   83.88 & 3.0 & 83;85 & 0.85 & 58880 & 14 \\
 G37.430+1.518    & 18 54 14.23 &    04 41 41.10 &   41.32 & 73.8 & 40;43 & 31.79 & 58862 & 8,11,14 \\
 G37.546$-$0.112  & 19 00 16.05 &    04 03 16.09 &   50.03 & 1.1 & 49;51 & 0.58 & 58866 & 14 \\
 G40.282$-$0.219  & 19 05 41.21 &    06 26 12.69 &   74.35 & 1.7 & 72;81 & 2.61 & 58716 & 14 \\
 G40.425+0.700    & 19 02 39.62 &    06 59 10.50 &    6.62 & 7.9 & 5;16 & 5.60 & 58827 & 14 \\
 G42.034+0.190    & 19 07 28.18 &    08 10 53.47 &   11.50 & 4.6 & 10;15 & 6.09 & 58872 & 14 \\
 G43.149+0.013    & 19 10 11.05 &    09 05 20.40 &   13.64 & 2.4 & 13;15 & 0.94 & 58766 & 1,9,11,14 \\
 G43.890$-$0.784  & 19 14 26.39 &    09 22 36.50 &   51.66 & 2.4 & 45;53 & 1.71 & 58805 & 11,14 \\
 G45.804$-$0.356  & 19 16 31.08 &    11 16 12.01 &   60.01 & 1.3 & 59;61 & 0.18 & 58762 & 14 \\
 G49.043$-$1.079  & 19 25 22.25 &    13 47 19.50 &   36.31 & 0.9 & 35;37 & 0.49 & 58721 & 14 \\
 G49.265+0.311    & 19 20 44.85 &    14 38 26.91 & $-$4.55 & 1.4 & $-$7;$-$3 & 1.83 & 58827 & 14 \\
 G49.349+0.413    & 19 20 32.44 &    14 45 45.44 &   68.14 & 2.0 & 66;70 & 1.24 & 58860 & 14 \\
 G49.416+0.326    & 19 20 59.21 &    14 46 49.60 &$-$10.18 & 0.8 & $-$10.8;$-$9.5 & 0.45 & 58726 & 11,14 \\
 G49.489$-$0.369  & 19 23 39.82 &    14 31 04.90 &   59.01 & 2.3 & 57;60 & 1.37 & 58855 & 14 \\
 G49.490$-$0.388  & 19 23 43.95 &    14 30 34.20 &   56.13 & 10.1 & 55;57 & 5.30 & 58855 & 1,5,6,9,11,12,14 \\
 G49.599$-$0.249  & 19 23 26.61 &    14 40 16.99 &   64.05 & 4.3 & 61;67 & 9.97 & 58706 & 11,14 \\
 G50.035+0.582\tablefootmark{a}    & 19 21 15.45 &    15 26 49.20 & $-$5.02 & 1.1 & $-$6;$-$3 & 0.41 & 58717 & \\
 G52.199+0.723    & 19 24 59.84 &    17 25 17.90 &    3.31 & 2.8 & 2;5 & 1.01 & 58760 & 14 \\
 G52.663$-$1.092  & 19 32 36.07 &    16 57 38.40 &   65.22 & 2.6 & 64;67 & 1.24 & 58870 & 11,14 \\
 G59.783+0.065    & 19 43 11.25 &    23 44 03.30 &   26.96 & 4.3 & 26;29 & 1.97 & 58870 & 9,11 \\
 G79.736+0.991    & 20 30 50.67 &    41 02 27.60 & $-$5.59 & 4.7 & $-$7;$-$4 & 3.04 & 58713 & 11 \\
 G85.410+0.003\tablefootmark{a}    & 20 54 13.68 &    44 54 07.60 &$-$29.74 & 5.7 & $-$33;$-$27 & 5.63 & 58870 & \\
 G107.298+5.639\tablefootmark{a}   & 22 21 22.50 &    63 51 13.00 & $-$7.34 & 25.1 & $-$9;$-$6 & 8.44 & 58677 & \\
 G109.871+2.114   & 22 56 17.90 &    62 01 49.65 & $-$4.07 & 44.5 & $-$5;$-$1 & 10.99 & 58902 & 4,5,11 \\
 G111.542+0.777   & 23 13 45.36 &    61 28 10.55 & $-$61.30 & 36.7 & $-$55;$-$62 & 49.46 & 58702 & 1,4,5,11 \\
 G133.947+1.064   & 02 27 03.82 &    61 52 25.40 &$-$44.54 & 601.3 & $-$47;$-$41 & 1334.22 & 58770 & 1,4,5,11 \\
 G183.348$-$0.575\tablefootmark{a} & 05 51 10.94 &    25 46 17.24 & $-$4.87 & 2.6 & $-$6;$-$4 & 1.15 & 58703 & \\
 G188.946+0.886   & 06 08 53.34 &    21 38 29.16 &   10.84 & 268.0 & 8;12 & 196.28 & 58892 & 3,4,5,6,9,11,13 \\
 G192.600$-$0.048 & 06 12 54.02 &    17 59 23.32 &   5.84 & 25.9 & 3;7 & 17.40 & 58723 & 12,13 \\
\hline

\end{tabular}
\tablebib{
(1) \citet{batrla1987}; (2) \citet{norris1987}; (3) \citet{kemball1988}; (4) \citet{koo1988}; (5) \citet{catarzi1993}; (6) \citet{caswell1993}; (7) \citet{macleod1993}; (8) \citet{gaylard1994}; (9) \citet{caswell1995b}; (10) \citet{moscadelli1996}; (11) \citet{blaszkiewicz2004}; (12) \citet{breen2010}; (13) \citet{breen2012b}; (14) \citet{breen2016}.
}
\tablefoot{\tablefoottext{a}{Newly detected 12.2\,GHz maser}}
\end{table*}

\subsection{Case of G107} \label{sec:g107_variability}
\begin{figure}
    \includegraphics[width=0.49\textwidth]{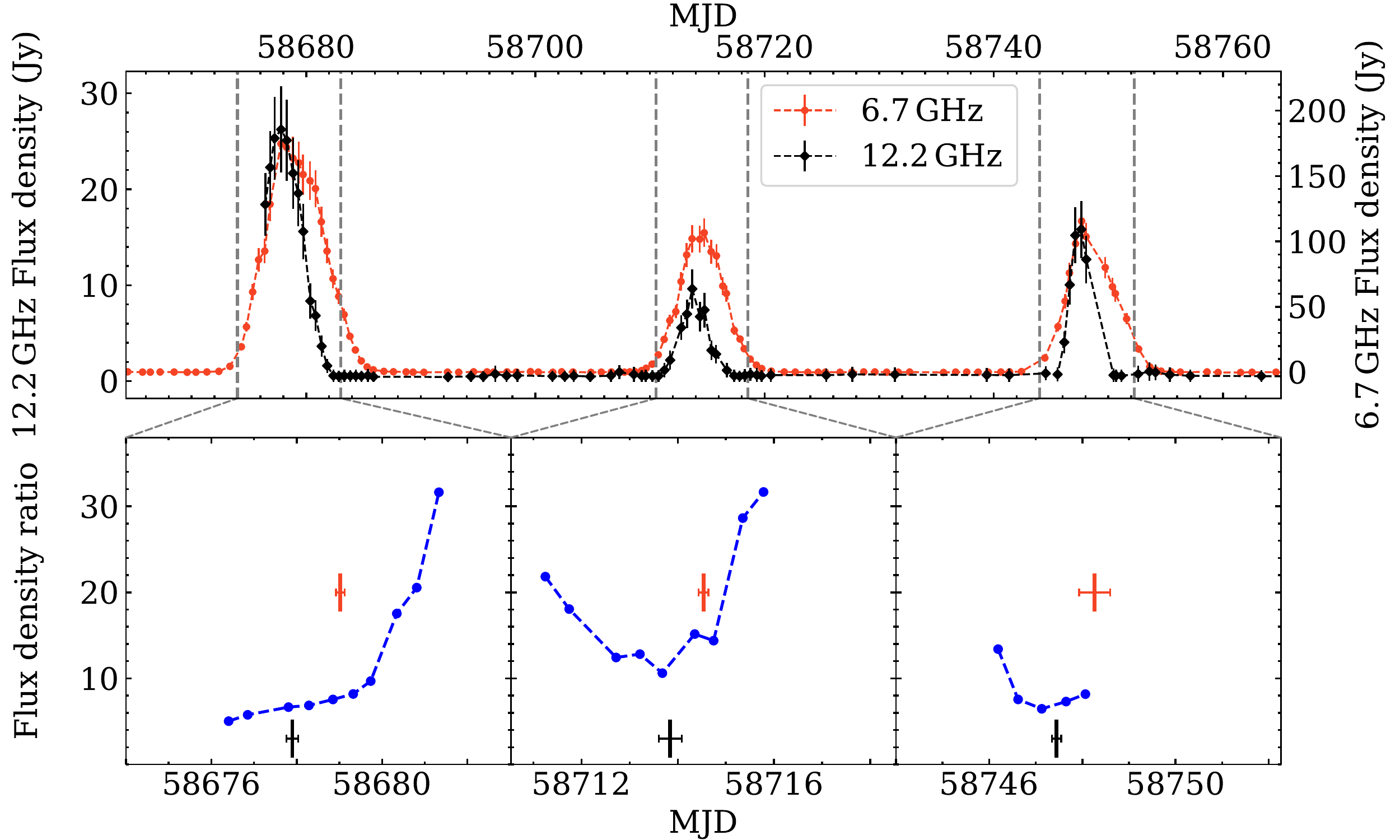}
    \caption{Light curves of the $-$7.4\kms\, methanol maser feature at 12.2\,GHz (black) and 6.7\,GHz (red) for G107.298+5.639 ({\it upper panel}). Temporal changes in the 6.7\,GHz to 12.2\,GHz flux density ratio are plotted (blue) ({\it lower panel}). The  thick vertical error bars denote the flare maxima at 12.2\,GHz (black) and 6.7\,GHz (red) calculated as the average value of the peak times obtained with the use of three methods (Table~\ref{tab:fit_res}), whereas the thin horizontal error bars mark the corresponding standard errors.
\label{fig:g107-lc}}
\end{figure}

The 12.2\,GHz maser emission follows periodic variations observed in the methanol 6.7\,GHz and hydroxyl (1.665/1.667\,GHz) maser lines \citep{szymczak2016,olech2020}. The upper panel in Fig.\,\ref{fig:g107-lc} presents the light curves of both methanol transitions. In order to estimate the parameters of flare profiles, we fitted an asymmetric power function \citep{david1996,szymczak2011}, Gaussian function, and second-order polynomial to the 6.7 and 12.2\,GHz data. The flare maximum and timescale of variability FWHM of the flare are listed in Table\,\ref{tab:fit_res} for each transition and three cycles. For the asymmetric function fitting we get, in addition, the ratio of rise to decay time of the flare, which ranges from 0.8 to 1.1 and 0.7 to 0.9 for the 12.2 and 6.7\,GHz lines, respectively. 
This suggests that the flare profile at 12.2\,GHz is slightly more symmetric than the one at 6.7\,GHz. The full width at half maximum (FWHM) values of 12.2\,GHz flares (1.8$-$4.5\,d) vary from cycle to cycle much more than those of 6.7\,GHz flares (4.4$-$5.8\,d), suggesting a greater variability of the 12.2\,GHz emission. 

There is a systematic delay between 6.7 and 12.2\,GHz flare peaks (Table\,\ref{tab:fit_res}) for each of the three cycles (lower panel in Fig.\,\ref{fig:g107-lc}). The average delay obtained from the results of the three fits is 0.9$\pm$0.3\,d. The VLBI maps indicate that the 6.7\,GHz emission near $-$7.35\kms\, comes from clouds located $\sim$150\,au from the putative position of the central star \citep{olech2020}. Thus, if the 6.7\,GHz and 12.2\,GHz maser coexist, then the observed delay cannot be explained by a simple geometrical effect. We may speculate that the actual size of the region conducive for 6.7\,GHz maser emission is more extended than it would be for 12.2\,GHz maser emission and both maser regions may not coincide precisely. Further interferometric studies are needed to verify this possibility.

The lower part of Figure\,\ref{fig:g107-lc} shows variations in the 6.7\,GHz to 12.2\,GHz flux density ratio ($R_{6/12}$) over the cycles. There are significant differences in the temporal behaviour of $R_{6/12}$ over the flare profile from cycle to cycle. For the best sampled observations of the second cycle, with a peak around MJD 58714, $R_{6/12}$ 
falls from 22 at flare onset to 10.4 at flare maximum and then increases to 32 as the flare decays.
The average value of $R_{6/12}$ for the three cycles around the flare peaks is 8.1 which is very close to the median ratio reported in \citet{caswell1995a} and \citet{breen2014}.

\begin{table}
\centering
\caption{Flare parameters retrieved from the three model fitting. The points with the flux density greater than 50\% of the peak value were used for the
parabola fitting. $S_\mathrm{p}$ is the fitted flux density at flare peak.
 \label{tab:fit_res}}
\begin{tabular}[c]{ccc|ccc}
\hline
\multicolumn{3}{c}{6.7\,GHz} & \multicolumn{3}{|c}{12.2\,GHz}\\
Epoch  & FWHM & $S_\mathrm{p}$ & Epoch  & FWHM & $S_\mathrm{p}$\\
(MJD) & (d) & (Jy) & (MJD) & (d) & (Jy)\\
\hline
\multicolumn{6}{c}{Asymmetric power function} \\
\hline
58678.80 & 5.76 & 175.5 & 58677.85 & 4.46 & 26.0\\
58714.46 & 4.41 & 107.6 & 58713.82 & 2.90 & 8.4\\
58748.20 & 4.52 & 112.0 & 58747.43 & 1.93 & 16.5\\
\hline
\multicolumn{6}{c}{Gaussian function} \\
\hline
58679.25 & 5.18 & 191.4 & 58677.93 & 3.81 & 27.2 \\
58714.63 & 4.37 & 109.5 & 58713.82 & 2.85 & 8.4 \\
58748.59 & 4.53 & 113.4 & 58747.46 & 1.81 & 16.9 \\
\hline
\multicolumn{6}{c}{2-nd order polynomial} \\
\hline
58679.02 & - & 162.6 & 58677.91 & -  & 25.6 \\
58714.51 & -  & 99.5  & 58713.90 & -  & 8.1 \\
58748.00 & -  & 110.2 & 58747.45 & -  & 16.2 \\
\hline
\end{tabular}
\end{table}

\section{Comments on previously known sources}
In this section, we provide commentary on our observational data with the aim of providing useful information on the 6.7 to 12.2\,GHz flux density ratio. Taken together with results of previous surveys given in Table \ref{tab:dec_list}, it is possible to estimate the degree and timescale of variability.

\textit{G30.225$-$0.180}. The 12.2\,GHz maser spectrum detected in September and December 1992 by \citet{caswell1995b} was composed of two features at  V$_{12.2}$ =  112.8 and 113.6\kms,\, with a peak flux density S$_{12.2}$ $\sim$ 2.8 and 1.2\,Jy, respectively. Our observations reveal the spectrum of different shape with a peak flux density of S$_{12.2}$ = 6.7\,Jy at V$_{12.2}$ = 113.3\kms. There was no emission found, with a 5$\sigma$ sensitivity level of 1.8\,Jy, in March 2010 \citep{breen2016}. This implies significant variability on timescales of 9-27\,yr. The emission at a velocity lower than 111\kms\, (Fig.\,\ref{fig:specrtas_nonewdec}) comes from another source, namely, G30.198$-$0.169 \citep{caswell1995b, breen2016}, which was detected in the beam sidelobe.

\textit{G32.744$-$0.075}. A comparison of our observations with those in the literature suggests that the overall spectral profile of the 12.2\,GHz maser emission has been unchanged over more than 40\,yr \citep{caswell1993}. The peak flux density of individual features differ by $\le$40\% on timescales of 9-27\,yr \citep{caswell1993,caswell1995b,breen2016}, suggesting only modest variability.

\textit{G33.641$-$0.228}. 
The brightest features in our spectrum have similar velocities to those reported in \citet{breen2016}, but their flux densities differ by a factor of 0.3-2.3. Furthermore, the emission at V$_{12.2}$ =  58.8\kms\, decreased by an order of magnitude as compared to that observed in 2002 by \citet{blaszkiewicz2004}. This implies significant variability on timescales of 9-18\,yr. The source also shows a variety of variability patterns in the 6.7\,GHz methanol maser emission from short (2-5\,d) flares, with a seven-fold rise within 24\,hr \citep{fujisawa2012,Fujisawa2014}, up to quasi-periodic ($>$500\,d) variations of low amplitude \citep{olech2019}.

\textit{G35.132$-$0.744}. The strongest feature at V$_{12.2}$ =  31.2\kms\, that was observed by \citet{breen2016} increased by a factor of 2, whereas the feature at V$_{12.2}$ = 29.7\kms\, increased by one order of magnitude. This implies strong variability on a timescale of 9\,yr.

\textit{G35.197$-$0.743}. The spectrum observed in April 1988 \citep{caswell1993} is complex, with a prominent feature at V$_{12.2}$ = 30.5\kms,\, with a peak flux of S$_{12.2}$ = 44\,Jy. The intensity of this feature decreased by $\sim$30\%, while the emission peak at V$_{12.2}$ = 28.5\kms\, increased by the same amount after $\sim$4.5\,yr \citep{caswell1995b}. \citet{macleod1993} and \citet{blaszkiewicz2004} reported the spectrum similar to that observed in December 1992 by \citet{caswell1995b},  while the \citet{breen2016} spectrum is similar to that presented in \citet{caswell1993}. Thus, from the data reported in the literature we infer a variability of 30-40\% on a timescale of 4-20\,yr.  Our survey suggests that this level of variability remains on timescale of $\ge$30\,yr for the redshifted emission (30.5\kms) but the feature at V$_{12.2}$ = 28.2\kms\, increased by a factor of 4 as compared to the spectrum from \citet{breen2016}. 

\textit{G35.200$-$1.736}. \citet{norris1987} reported the spectrum, consisting of two features at V$_{12.2}$ = 41.5 and 45.0\kms\,with peak flux densities of S$_{12.2}$ = 100 and 146\,Jy, respectively. A similar spectrum was seen after 1-5 yr  with slight ($\le$15\%) decreases of peak intensity \citep{caswell1993, caswell1995b,catarzi1993}. Observations in 2002 \citep{blaszkiewicz2004} and 2010 \citep{breen2016} displayed an ongoing complex spectrum with two prominent features but with the flux density of the strongest feature at V$_{12.2}$ = 45.0\,\kms\, decreasing to S$_{12.2}$ = $\sim$80\,Jy in 2002 and S$_{12.2}$ = 46\,Jy in 2010. 
We report only a single feature with a peak of S$_{12.2}$ = 12\,Jy at V$_{12.2}$ = 45.2\kms. We conclude that the 12.2\,GHz line intensity declined by an order of magnitude within 32\,yr. 

\textit{G37.430+1.518}. Our spectrum, with a single feature of S$_{12.2}$ = 74\,Jy at V$_{12.2}$ = 41.3\kms, is very similar to that reported by \citet{blaszkiewicz2004} and \citet{breen2016}, implying a variability lower than 20\% on a timescale of 17\,yr. The emission observed 26\,years ago \citep{gaylard1994} was a factor of 6 weaker, suggesting significant variability on longer timescales.

\textit{G40.282$-$0.219}. The three-feature spectrum is similar to that observed nine\,years ago \citep{breen2016} but the flux density of the primary feature at V$_{12.2}$ = 74.3\kms\, decreased by $\sim$50\%.

\textit{G42.034+0.190}. We detected a weak and complex 12.2\,GHz spectrum of similar shape to that shown in \citet{breen2016}. In general, the intensity of features increased by a factor of 1.8 to 4.5 after nine\,years.

\textit{G43.149+0.013}. The 12.2\,GHz spectrum consisted of a single feature at V$_{12.2}$ = 13.6\kms. Its intensity declined by 35\% over 26\,yr \citep{caswell1995b, breen2016}.

\textit{G43.890$-$0.784}. Little variability is visible in the two features at V$_{12.2}$ = 47.4 and 51.8\kms\, when compared to spectra from 2002 \citep{blaszkiewicz2004} an 2010 \citep{breen2016}.
From our observations, it can be seen that the emission at V$_{12.2}$ = 51.8\kms\, decreased by a factor of 2. There was no emission detected at V$_{12.2}$ = 47.7\kms, whereas a new feature appeared at V$_{12.2}$ = 45.3\kms, indicating significant variability. 

\textit{G49.043$-$1.079}. 
 Nine years ago, this 12.2\,GHz source had a complex spectra \citep{breen2016}. In our observations, the strongest feature they detected (S$_{12.2}$ = 7.1\,Jy at V$_{12.2}$ = 37.4\kms) is now only $\sim$1\,Jy.
This suggests strong variability on this timescale. 

\textit{G49.416+0.326}. The spectrum is similar to that presented in \citet{breen2016}, with little variability above the noise level.

\textit{G49.490$-$0.388}. The spectrum is composed of the emission from \textit{G49.489$-$0.369} and \textit{G49.490$-$0.388} as part of the W51 complex. \citet{batrla1987} discovered a single feature, with a peak flux density of  S$_{12.2}$ = 8\,Jy at V$_{12.2}$ $\sim$56\kms, and a broad  absorption feature (from 60 to 75\kms). \citet{catarzi1993} observed a similar spectrum with a peak flux density of S$_{12.2}$ = 20\,Jy without absorption. \citet{caswell1993,caswell1995b} also reported one feature with the peak flux density of S$_{12.2}$ = 14 and 21\,Jy, respectively with barely visible absorption. A simple spectrum with a single feature at V$_{12.2}$ = 56.2\kms\, with S$_{12.2}$ = 13.8\,Jy was also observed by \citet{blaszkiewicz2004}. The flux density of this feature dropped to 4.5-5.5\,Jy (in 2008 and 2010, \citealt{breen2016}) and increased to $\sim$10\,Jy during our observations. This suggests that the emission towards G49.490$-$0.388 shows significant and complicated temporal changes. Features at velocity $>$56.6\kms\, were also visible only in 2010 \citep{breen2016} that might suggest variability of G49.489$-$0.369. It is difficult, given our beam size, to distinguish unequivocally which spectral feature comes from which object; hence, the velocity ranges marked in Fig.\,\ref{fig:specrtas_nonewdec} are taken from \citet{breen2016}. 

\textit{G49.599$-$0.249}. The spectrum is almost the same as the one seen in \citet{breen2016}, with the exception of the 63.0\kms\, feature, which decreased by a factor of 2 after 9 yr. 

\textit{G59.783+0.065}. The feature at V$_{12.2}$ = 27.1\kms\, declined from S$_{12.2}$ = 15.8\,Jy in 1992 \citep{caswell1995b} to S$_{12.2}$ = 9.2\,Jy in 2002 \citep{blaszkiewicz2004} and S$_{12.2}$ = 4.3\,Jy during our observations. The second feature at V$_{12.2}$ = 17.0\kms\,, with a peak flux of S$_{12.2}$ = 4\,Jy \citep{caswell1995b} increased by $\sim$50\% \citep{blaszkiewicz2004} and decreased below our sensitivity limit of $\sim$1\,Jy. 

\textit{G109.871+2.114}. 
Observations made in 1990 \citep{catarzi1993}, with a peak flux density of S$_{12.2}$ = 128.8\,Jy at V$_{12.2}$ = $-$4.15\kms\,, 
found a similar spectral profile shape to that seen in 1987 \citep{koo1988}, but with flux density  that was a factor of $\sim$2 lower. \citet{blaszkiewicz2004} revealed a similar profile but with peak flux densities that were a factor of 3 lower than \citet{koo1988}. We detected a complex spectrum with the emission from V$_{12.2}$ = $-$5 to $-$1\kms\, and a peak flux density of S$_{12.2}$ = 45\,Jy at  V$_{12.2}$ = $-$4.1\kms. Thus, the source is significantly variable, namely,  by  a factor of 7 on a timescale of $\sim$30\,yr.

\textit{G111.542+0.777}. 
\citet{batrla1987} discovered a double peaked spectrum, the brightest feature at V$_{12.2}$ = $-$56.3\kms\,, with S$_{12.2}$ $\sim$200\,Jy
remained stable within 15\% over 1-3\,yr \citep{koo1988,catarzi1993}. The spectral shape was found to be unchanged in 2002 \citep{blaszkiewicz2004} but the peak flux density decreased by a factor of 2. Here, we present a very different spectrum with several blended features from $-$62 to $-$55\kms. The feature at V$_{12.2}$ = $-$56.3\kms\, dimmed by a factor of 20 compared to that reported by \citet{batrla1987}, whereas the feature near V$_{12.2}$ = $-$61.5\kms\, increased by a factor of less than 2. This suggests strong variability on a timescale of $\ge$20\,yr. 

\textit{G188.946+0.886}.
The flux density of the strongest feature at V$_{12.2}$ = 10.4\kms\  increased by $\sim$30\% between 1987 and 1992 \citep{koo1988,kemball1988,caswell1993,catarzi1993,caswell1995b}. The profile of the spectrum significantly transformed in 2008 \citep{breen2012b} but the intensity remained unchanged within 10\%. We found spectra similar to those reported in \citet{breen2012b}, and the peak flux of feature at V$_{12.2}$ = 10.8\kms\, increased by $\sim$15\%. 
More information on the variability of this object is provided in Sect.~\ref{sec:obser}.

\textit{G192.600$-$0.048}. A faint (0.5-0.6\,Jy) feature at V$_{12.2}$ = 3.6\kms\, was detected in June and December 2008 \citep{breen2012b}. Our observations revealed a complex spectrum with the strongest feature of S$_{12.2}$ = 26\,Jy at V$_{12.2}$ = 5.8\kms. Significant change in the spectral profile and intensity is likely related to a significant outburst of a 6.7\,GHz maser that occurred in mid-2015 \citep{moscadelli2017,szymczak2018}. 

There is a group of 12.2\,GHz masers that have gone undetected in the present survey (Table\,\ref{tab:non-det}) and 13 out of those 17 masers were observed with a 5$\sigma$ sensitivity of 0.7 to 4.2\,Jy, that is, lower than the peak flux densities at the time of their detection from the literature. We failed to detect, with 5$\sigma$ sensitivity of 0.6-0.9\,Jy, the following objects: G42.698$-$0.147, G45.467+0.053, G94.602$-$1.796, and G196.454$-$1.677, with a peak flux density of 1.2 to 12.3\,Jy in earlier observations \citep{caswell1995b,blaszkiewicz2004,breen2012b,breen2016}.
These four sources may be variable on timescales smaller than 9-28\,yr.

\section{Discussion}
\subsection{Detection rate}
The present observations of our sample of 153 6.7\,GHz masers resulted in the detection of 36 12.2\,GHz methanol maser sources, corresponding to a $\sim$24\% detection rate. This is about a factor of 2 lower than the overall detection rate of 43\% in the Galactic longitude range of 290\degr\,$\le l \le$ 60\degr\, \citep{breen2016}. We compared our detection rate with the rates from \citet{breen2016} as a function of 10\degr\, longitude bins. For the 30\degr$-$40\degr, 40\degr$-$50\degr\,, and 50\degr$-$60\degr\, longitude bins, our detection rates are lower by 20\%, 9\%, and 2\%, respectively, than those reported in \citet{breen2016}. This confirms a significant lowering of our sensitivity for low declination sources as shown in Table\,\ref{tab:non-det}. The reason for this is a degradation of sensitivity due to interference from geostationary satellites or imperfect removal of some corrupted scans during the edition of spectra. The detection rate in the range of 70\degr\,$\le l \le$ 90\degr\, decreases to
$\sim$16\% and shows a statistically significant difference from that for the neighbouring range of 50\degr$-$60\degr\, (23.5\%). 

\subsection{Flux density ratio}
Our survey indicates that the flux density of individual features and integrated flux density of the 6.7\,GHz maser sources are all greater than those of their 12.2\,GHz counterparts. This is consistent with conclusions inferred from more comprehensive statistical analysis based on a large sample \citep{breen2011}. In the following, we confine discussion to the flux density ratio $R_{6/12}$ for the spectral features with the same velocity in both transitions contemporaneously observed in order to exclude the possible effect of variability. There is strong observational evidence of spatial coincidence of the 6.7 and 12.2\,GHz masers to within a few milliarcseconds, especially when the spectral profiles of both  transitions are similar \citep{menten1992,norris1993,minier2000,moscadelli2002}. Thus, it is suggestive that the flux density comparison can be meaningful in the absence of high angular resolution maps. We found 70\% of the detected 12.2 GHz methanol maser peaks are coincident in velocity with the 6.7 GHz maser peak.

\begin{figure}
    \includegraphics[width=0.48\textwidth]{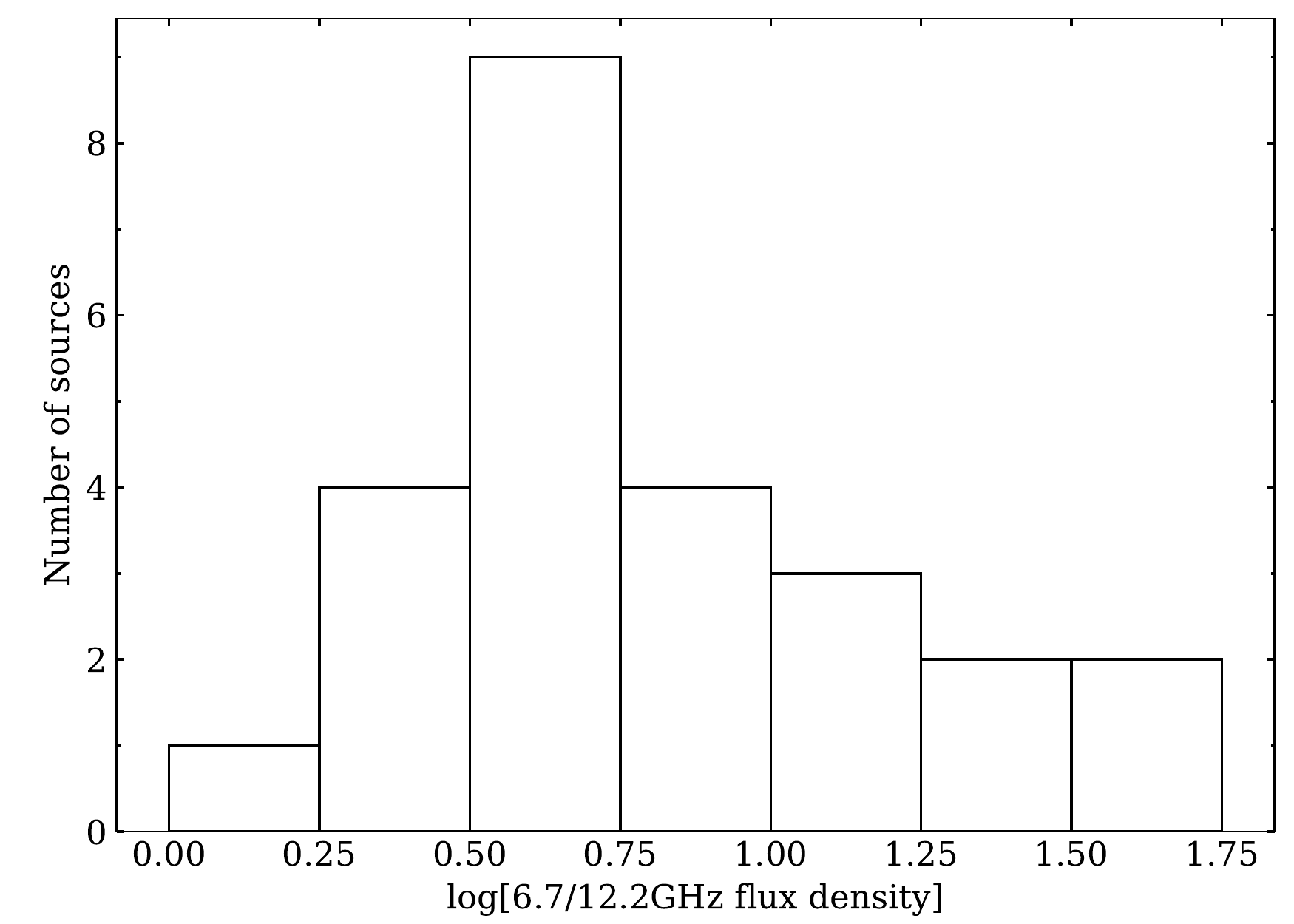}
    \caption{Histogram of the 6.7\,GHz to 12.2\,GHz peak flux density ratio. 
    \label{fig:histo-ratio}}
\end{figure}

\begin{table*}
\centering
\caption{6.7\,GHz to 12.2\,GHz flux density ratio ($R_{6/12}$) for the Gaussian fitted features for all the sources contemporaneously observed at both maser transitions. The fitted peak velocity ($V$), peak flux density at 6.7\,GHz ($S_{6.7}$) and at 12.2\,GHz ($S_{12.2}$) are listed. The values in brackets correspond to uncertainty, derived from covariance matrix of the least-squares fit method. 
\label{tab:ratios_oldsrc}}
\begin{tabular}[c]{crrrr}
\hline
Source (l   b) & $V$\hspace{14pt} & $S_{6.7}$\hspace{14pt} & $S_{12.2}$\hspace{10pt} & $R_{6/12}$\hspace{10pt} \\
(\degr \hspace{0.5cm} \degr)  & (km\,s$^{-1}$) & (Jy)\hspace{14pt} & (Jy)\hspace{12pt} & \\
\hline
G35.200$-$1.736 & 44.62    & 34.3(0.4)  &  2.1(0.4)  & 16.0(2.4)\\
                & 45.21    & 17.8(0.2)  & 11.9(0.3)  &  1.5(0.1)\\
G49.416+0.326   & $-$10.13 &  2.3(0.1)  &  0.8(0.2)  &  2.7(0.6)\\
G49.599$-$0.249 & 65.51    & 10.5(2.0)  &  3.0(0.6)  &  3.5(1.0)\\
                & 62.93    & 15.9(0.1)  &  3.7(0.1)  &  4.3(0.1)\\
                & 64.14    & 15.4(0.3)  &  4.5(0.2)  &  3.4(0.2)\\
                & 64.61    & 16.6(2.3)  &  4.2(0.2)  &  3.9(0.6)\\
                & 65.34    & 15.3(4.0)  &  3.0(0.6)  &  5.1(1.7)\\
G85.410+0.003   & $-$31.55 &  9.9(3.5)  &  1.2(0.2)  &  8.4(3.1)\\
                & $-$28.65 &  5.7(0.2)  &  2.5(0.2)  &  2.3(0.2)\\
G107.29+5.64    & $-$7.35  &170.2(1.0)  & 25.8(0.4)  &  6.6(0.2)\\
G109.871+2.114  & $-$3.72  &125.7(22.0) &  3.8(0.2)  & 32.7(5.9)\\
                & $-$1.80  &245.8(1.4)  &  4.8(0.2)  & 50.9(1.6)\\
G111.542+0.777  & $-$61.33 &154.9(0.6)  & 36.7(0.4)  &  4.2(0.1)\\
                & $-$60.77 &118.7(0.5)  & 14.6(0.3)  &  8.1(0.2)\\
                & $-$58.08 &230.2(3.2)  & 12.0(0.3)  & 19.2(0.6)\\
                & $-$57.60 &151.2(3.5)  &  7.0(0.8)  & 21.6(2.3)\\
                & $-$57.22 & 91.7(6.0)  &  6.6(0.5)  & 13.9(1.3)\\
                & $-$56.75 & 93.9(0.7)  & 13.4(0.3)  &  7.0(0.2)\\
G188.946+0.886  & 10.43    &400.4(3.9)  &109.2(3.2)  &  3.7(0.2)\\
                & 10.86    &858.2(4.7)  &269.6(1.8)  &  3.2(0.1)\\
G192.600$-$0.048&  3.93    &  8.8(0.4)  &  4.0(0.2)  &  2.2(0.2)\\
                &  4.79    & 13.5(0.4)  &  5.3(0.3)  &  2.5(0.2)\\
                &  5.46    & 55.6(1.4)  &  5.2(0.5)  & 10.7(1.0)\\
                &  5.83    &139.7(3.3)  & 25.5(0.5)  &  5.5(0.2)\\
\hline
\end{tabular}
\end{table*}

For the 6.7 and 12.2\,GHz spectra of nine sources taken within 7\,days of each other, we performed Gaussian function fits to obtain the velocity and flux density of feature peaks. For all features with the same peak velocity ($\pm0.1$\kms) at both transitions, the flux density ratio, $R_{6/12}$, was determined (Table\,\ref{tab:ratios_oldsrc}). The range of $R_{6/12}$ is 1.5 to 50.9, with a median value of 5.1 (Fig.\,\ref{fig:histo-ratio}). 
Our median is comparable to that determined by \citet{caswell1995a}. 
The analysis by \citet{breen2011} of the statistical properties of 580 southern sources found a median peak-to-peak ratio of 4.3. Although their samples are more numerous, with little overlap with ours, the median ratios are similar. 
This suggests that each of these samples come from a similar population of HMYSOs. 

There is a significant dispersion of $R_{6/12}$ for spectral features of some sources (e.g. G111.524+0.777, Table\,\ref{tab:ratios_oldsrc}). Furthermore, our observations of G107 reveal that around the flare maxima, $R_{6/12}$ varies by up to 50\% between two consecutive cycles. In addition, the temporal behaviour of $R_{6/12}$ is poorly repeated from cycle to cycle, even though a general trend remains, that is, the ratio reaches a minimum around the flare peak and is larger at the onset and final stages of the flare profile. This could be caused by variations in the physical conditions along the maser path length of $10^{16}$\,cm \citep{moscadelli2002} for individual features on different timescales. 

In G107, the maser intensity varies with a period of 34.4\,days, perhaps due to changes of infrared emission \citep{olech2020}; it is difficult to identify processes that can significantly change the gas density, molecule abundance, and kinetic temperature in the maser regions on such a short timescale. Since the heating and cooling times of dust grains are less than a few minutes and one day, respectively, for the optically thick case \citep{vanderWalt2009, johnstone2013}, rapid changes in the maser intensity may be related to variations in the dust temperature ($T_{\mathrm{d}}$). Timescales of gas heating and cooling are 2-3 orders of magnitude greater than those of dust grains (e.g. \citealt{johnstone2013}); thus, the gas temperature variation can affect the maser emission in G107 on timescales of several months. 

Numerical models of infrared pumping indicate that the 6.7 and 12.2\,GHz methanol masers over a wide range of the gas densities and kinetic temperatures \citep{cragg2002,cragg2005}, but there is a narrow range of dust temperature, $T_{\mathrm{d}}$, for which the flux ratio varies quite rapidly (\citealt{cragg2005}, their Fig.2). When the dust temperature increases from $\sim$130\,K to $\sim$170\,K, $R_{6/12}$ decreases by a factor of 2 and vice versa. Therefore, the variations in $T_{\mathrm{d}}$ qualitatively explain the observed change in the flux ratio during the flare of G107. However, it should be noted that the temporal behaviour of $R_{6/12}$ varies considerably from cycle to cycle, which is possibly the result of small variations in the physical conditions along the maser path and the degree of saturation (e.g. \citealt{breen2012a}). High-cadence monitoring observations combined with a high angular resolution are required to resolve this discrepancy.

\subsection{Variability}
\begin{figure}
\includegraphics[width=0.5\textwidth]{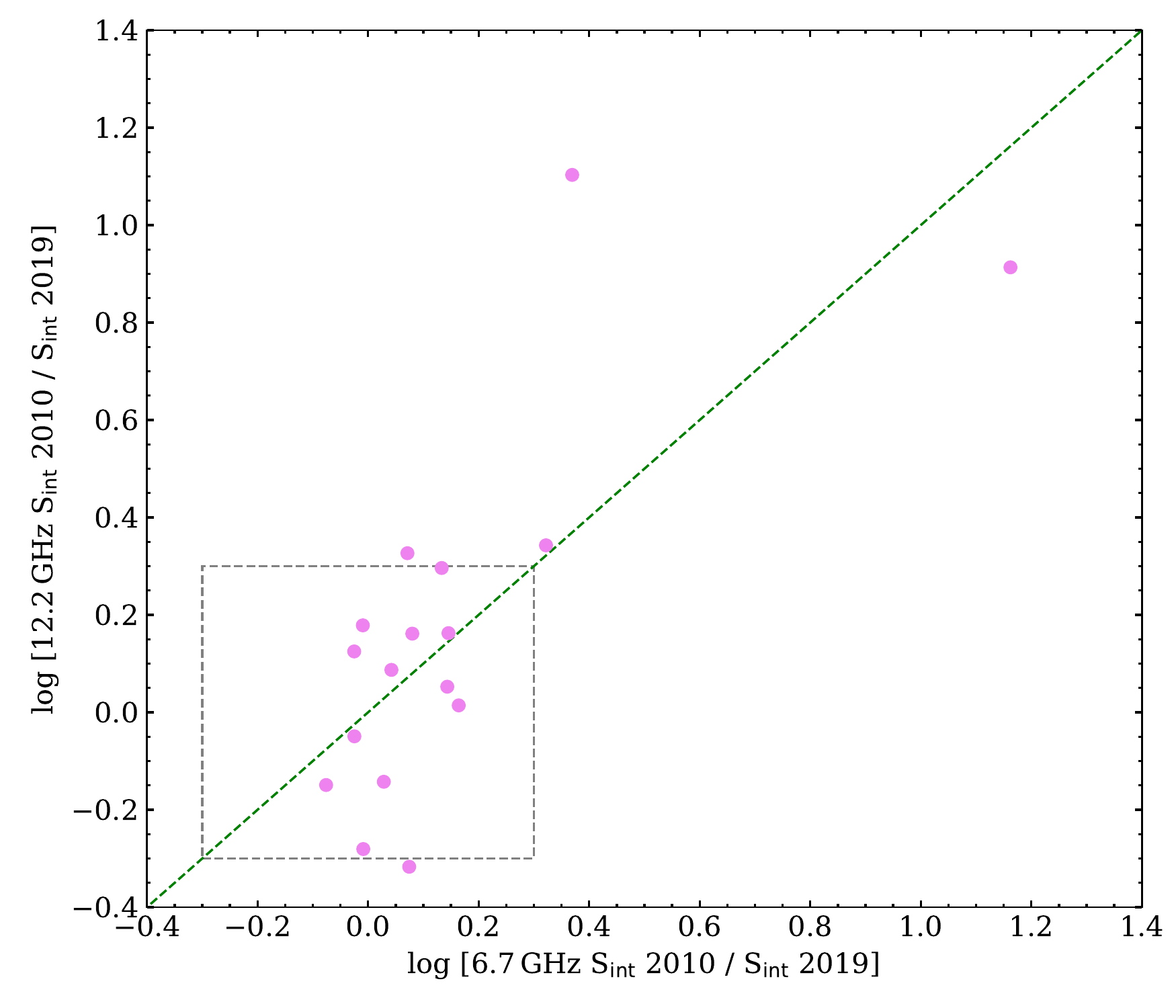}
\caption{ Relative change in the integrated flux density, $S_{\mathrm{int}}$ at 6.7 and 12.2\,GHz between 2010 \citep{breen2015,breen2016} and 2019 (this survey). The square marks 50\% level of variability. 
\label{fig:breenvstmsclog} } 
\end{figure}

\begin{figure}
\includegraphics[width=0.5\textwidth]{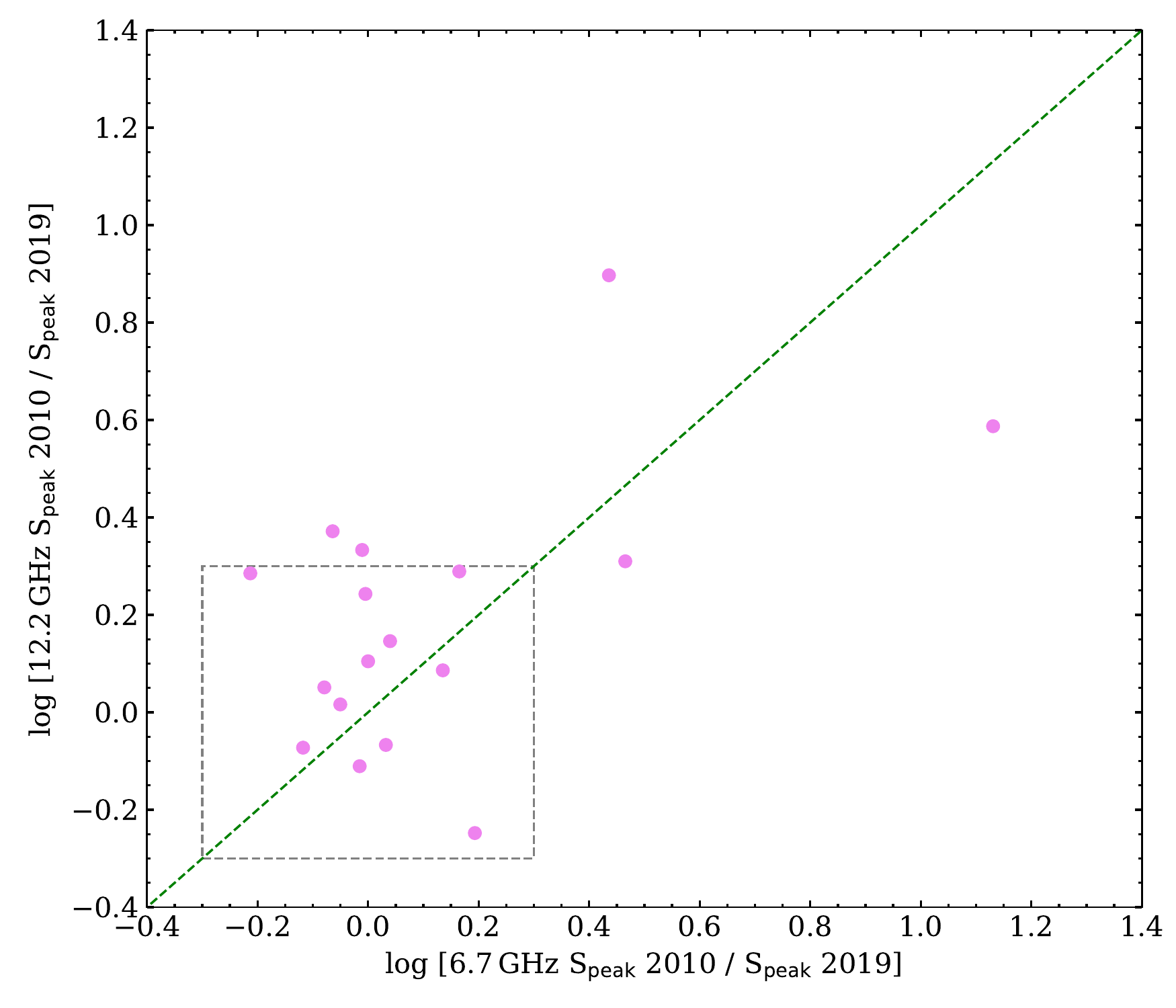}
\caption{ Relative change in the peak flux density, $S_{\mathrm{peak}}$ at 6.7 and 12.2\,GHz between 2010 \citep{breen2015,breen2016} and 2019 (this survey). The square marks 50\% level of variability. Data used to create this graph is presented in table \ref{tab:peak_comptab}.
\label{fig:breenvstmscpeak} }
\end{figure}

There are 24 detected sources
in our sample that are in common with those observed by \citet{breen2016} at 12.2\,GHz and \citet{breen2015} at 6.7\,GHz, 
providing an opportunity to study variations on a timescale of $\sim$10\,yr.
Seven objects with spectra comprised of multiple sources were not considered. In Fig.~\ref{fig:breenvstmsclog}, we present the relative changes in the integrated flux density ($S_{\mathrm{int}}$) between the two methanol maser transitions. In Fig.~\ref{fig:breenvstmscpeak}, the same analysis is presented, but for peak flux densities (Table \ref{tab:peak_comptab} shows exact values). Among 17 sources 
detected in two epochs, only two show $S_{\mathrm{int}}$ variability of more than 50\% in both transitions, these are: G35.200$-$1.736 and G49.043$-$1.079. The first object dimmed by a factor of 14.5 and 8.2 at 12.2\,GHz and 6.7\,GHz, respectively. The second object weakened by a factor of 12.7 at 12.2\,GHz but only a factor of 2.3 at 6.7\,GHz. G36.115+0.552 decreased by a factor of 2.2 at both lines. Two other objects (G45.804$-$0.356 and G49.265+0.311) varied by slightly more than 50\% at 12.2\,GHz but did not show significant variability at 6.7\,GHz. For the subsample of 17 sources, the median value of relative changes in $S_{\mathrm{int}}$ is 1.19 and 1.33 for the 6.7 and 12.2\,GHz lines, respectively. This general trend of a larger level of variability at 12.2\,GHz than that at 6.7\,GHz is more pronounced for the entire sample (Fig.\,\ref{fig:breenvstmsclog}, \ref{fig:breenvstmscpeak}) and is consistent with the standard model of methanol masers \citep{cragg2002,cragg2005}. They demonstrated that both masers operate in a wide range of gas density ($10^4$-$10^8$\,cm$^{-3}$), gas temperature (30-200\,K), and dust temperature (130-350\,K) but we can see that the slopes of the intensity versus these parameters are more flat at 6.7\,GHz than at 12.2\,GHz \citep{cragg2005}. Thus, long-term (9-10\,yr) changes of the parameters may lead to higher variability in the 12.2\,GHz line. Since high angular resolution interferometric studies have revealed that the 6.7 and 12.2\,GHz emission is spatially coincident, the effect of turbulence or changes in velocity coherence should be the same for both transitions. Table \ref{tab:peak_comptab} also presents comparison of the number of the features visible in 12.2\,GHz spectra among the literature data \citep{breen2016} and this survey. 
Many of the features generally remain unchanged ($\pm$1 feature), with the exception of
G35.200$-$1.736, for which a significant decrease of luminosity is observed in both 6.7 and 12.2\,GHz lines.

\section{Conclusions}
We report that we detected 36 12.2\,GHz methanol masers in our sample of 153, of which 4 are new detections, corresponding to detection rate of 24\%

Values of the 6.7\,GHz to 12.2\,GHz flux density ratio for spectral features at the same velocity, when both transitions are observed contemporaneously, are within the range from 1.5 to 51. The median value of 5.1 is similar to that reported for other large samples of HYMSOs. This ratio in the periodic source G107.298+5.639 is smallest when the flare reaches its maximum, but it varies considerably from cycle to cycle. 
It decreases from its maximum value at the onset through its minimum at the peak of the flare and then increases during the decay phase of the flare. This temporal behaviour appears to be consistent with the standard model of methanol masers when the dust temperature varies in the narrow range of 130-170\,K.

A minority (14\%) of the objects that we observed show strong ($>$50\%) variability at 6.7\,GHz over a timescale of 9-10\,yr, but at 12.2\,GHz, nearly half of the observed sources experienced strong variability. 
These results appear to be compatible with maser model predictions.

\section{Acknowledgements}
We thank the staff of the 32\,m telescope for assistance with the observations. 
We also thank the referee for the careful reading of the manuscript and recommendations.
The research has made use of the SIMBAD data base, operated at CDS (Strasbourg, France), as well as NASA's Astrophysics Data System Bibliographic Services. 
The 32\,m telescope is operated by the Institute of Astronomy, Nicolaus Copernicus University and supported by the Polish Ministry of Science and Higher Education SpUB grant.

\bibliography{librarian}
\bibliographystyle{aa}

\appendix
\section{Additional material}

\begin{table*}
\centering
\caption{List of targets towards which no 12.2\,GHz emission was detected in the survey. The source name in bold denotes previously detected masers.\label{tab:non-det}}
\begin{tabular}[c]{cccccc}
\hline
Name (l   b)  &  RA(J2000) & Dec(J2000)  & 5$\sigma$ & Epoch & Reference \\
(\degr \hspace{0.5cm} \degr) & (h \hspace{0.2cm}  m \hspace{0.2cm}   s) & (\degr \hspace{0.2cm} \arcmin  \hspace{0.2cm} \arcsec) & Jy & MJD & \\
\hline
G32.105$-$0.074 & 18 50 11.58 & $-$00 46 12.32 & 3.00 & 58827 &      \\
{\bf G33.133$-$0.092} & 18 52 07.82 & 00 08 12.80 & 3.80 & 58812 & 1, 3, 4\\
G33.199+0.001 & 18 51 55.34 & 00 14 19.38 & 4.90 & 58716 &      \\
G33.204$-$0.010 & 18 51 58.14 & 00 14 13.61 & 3.65 & 58813 &      \\
{\bf G33.317$-$0.360} & 18 53 25.30 & 00 10 43.90 & 4.20 & 58861 & 4\\
{\bf G33.393+0.010} & 18 52 14.62 & 00 24 52.90 & 3.00 & 58866 & 4\\
G33.725$-$0.120 & 18 53 18.78 & 00 39 05.00 & 4.85 & 58874 &      \\
G33.980$-$0.019 & 18 53 25.01 & 00 55 25.98 & 1.25 & 58812 &      \\
G34.096+0.018 & 18 53 29.94 & 01 02 39.40 & 4.30 & 58831 &      \\
{\bf G34.244+0.133} & 18 53 21.44 & 01 13 44.40 & 2.45 & 58859 & 1, 3, 4\\
{\bf G34.257+0.153} & 18 53 18.63 & 01 14 57.40 & 6.80 & 58866 & 1, 4\\
G34.267$-$0.210 & 18 54 37.25 & 01 05 33.70 & 2.75 & 58856 &      \\
G34.396+0.222 & 18 53 19.08 & 01 24 13.80 & 4.15 & 58860 &      \\
G34.411+0.235 & 18 53 17.99 & 01 25 25.26 & 3.00 & 58713 &      \\
G34.751$-$0.093 & 18 55 05.22 & 01 34 36.26 & 1.95 & 58860 &      \\
G34.757+0.025 & 18 54 40.74 & 01 38 06.40 & 2.30 & 58864 &      \\
G34.822+0.352 & 18 53 37.84 & 01 50 33.00 & 3.20 & 58875 &      \\
{\bf G35.025+0.350} & 18 54 00.66 & 02 01 19.30 & 2.35 & 58725 & 3, 4\\
G35.226$-$0.354 & 18 56 53.15 & 01 52 46.89 & 1.75 & 58706 &      \\
G35.247$-$0.237 & 18 56 30.38 & 01 57 08.88 & 3.55 & 58819 &      \\
G35.397+0.025 & 18 55 50.78 & 02 12 19.10 & 1.85 & 58868 &      \\
G35.417$-$0.284 & 18 56 59.02 & 02 04 55.65 & 2.00 & 58868 &      \\
G35.457$-$0.179 & 18 56 40.98 & 02 09 57.16 & 2.75 & 58875 &      \\
G35.588+0.060 & 18 56 04.22 & 02 23 28.30 & 5.00 & 58880 &      \\
{\bf G35.793$-$0.175} & 18 57 16.89 & 02 27 57.91 & 1.70 & 58882 & 4\\
G36.705+0.096 & 18 57 59.12 & 03 24 06.11 & 2.10 & 58883 &      \\
G36.918+0.483 & 18 56 59.78 & 03 46 03.60 & 1.50 & 58884 &      \\
G37.030$-$0.039 & 18 59 03.64 & 03 37 45.09 & 4.35 & 58894 &      \\
G37.554+0.201 & 18 59 09.98 & 04 12 15.54 & 2.75 & 58874 &      \\
G37.598+0.425 & 18 58 26.79 & 04 20 45.46 & 2.35 & 58867 &      \\
G37.735$-$0.112 & 19 00 36.84 & 04 13 20.00 & 2.05 & 58859 &      \\
G37.753$-$0.189 & 19 00 55.42 & 04 12 12.56 & 1.00 & 58802 &      \\
G37.767$-$0.214 & 19 01 02.27 & 04 12 16.60 & 1.55 & 58731 &      \\
G38.038$-$0.300 & 19 01 50.46 & 04 24 18.96 & 1.80 & 58882 &      \\
{\bf G38.119$-$0.229} & 19 01 44.15 & 04 30 37.42 & 4.25 & 58873 & 4\\
G38.203$-$0.067 & 19 01 18.73 & 04 39 34.29 & 1.95 & 58856 &      \\
G38.255$-$0.200 & 19 01 52.95 & 04 38 39.47 & 1.15 & 58762 &      \\
G38.258$-$0.073 & 19 01 26.25 & 04 42 19.90 & 1.80 & 58867 &      \\
G38.565+0.538 & 18 59 49.13 & 05 15 28.90 & 1.05 & 58810 &      \\
G38.598$-$0.212 & 19 02 33.46 & 04 56 36.40 & 1.75 & 58857 &      \\
G38.653+0.088 & 19 01 35.24 & 05 07 47.36 & 1.80 & 58857 &      \\
G38.916$-$0.353 & 19 03 38.65 & 05 09 42.49 & 1.40 & 58818 &      \\
G39.388$-$0.141 & 19 03 45.31 & 05 40 42.68 & 1.05 & 58712 &      \\
G40.623$-$0.138 & 19 06 01.63 & 06 46 36.50 & 0.90 & 58714 &      \\
G40.934$-$0.041 & 19 06 15.37 & 07 05 54.49 & 1.80 & 58819 &      \\
G41.121$-$0.107 & 19 06 50.24 & 07 14 01.49 & 1.70 & 58867 &      \\
G41.123$-$0.220 & 19 07 14.85 & 07 11 00.69 & 1.40 & 58860 &      \\
G41.156$-$0.201 & 19 07 14.37 & 07 13 18.10 & 0.90 & 58713 &      \\
G41.226$-$0.197 & 19 07 21.37 & 07 17 08.17 & 0.85 & 58704 &      \\
{\bf G41.347$-$0.136} & 19 07 21.84 & 07 25 17.27 & 1.80 & 58856 & 4\\
G42.13+0.52 & 19 06 28.90 & 08 25 10.00 & 4.05 & 58705 &      \\
G42.303$-$0.299 & 19 09 43.59 & 08 11 41.41 & 1.65 & 58705 &      \\
G42.435$-$0.260 & 19 09 49.85 & 08 19 45.40 & 0.85 & 58705 &      \\
{\bf G42.698$-$0.147} & 19 09 55.06 & 08 36 53.45 & 0.85 & 58718 & 4\\
G43.038$-$0.453 & 19 11 38.98 & 08 46 30.71 & 1.05 & 58722 &      \\
G43.074$-$0.077 & 19 10 22.05 & 08 58 51.49 & 3.50 & 58832 &      \\
G43.180$-$0.518 & 19 12 09.02 & 08 52 14.30 & 0.95 & 58766 &      \\
G43.796$-$0.127 & 19 11 53.97 & 09 35 53.50 & 0.95 & 58726 &      \\
\hline
\end{tabular}
\end{table*}

\begin{table*}
\centering
\begin{tabular}[c]{cccccc}
\multicolumn{6}{c}{Table \ref{tab:non-det} continued.}\\
\hline
Source (l   b)  &  Ra & Dec  & 5$\sigma$ & Epoch & Reference \\
(\degr \hspace{0.5cm} \degr) & (h \hspace{0.2cm}  m \hspace{0.2cm}   s) & (\degr \hspace{0.2cm} \arcmin  \hspace{0.2cm} \arcsec) & Jy & MJD & \\
\hline
G44.310+0.041 & 19 12 15.81 & 10 07 53.52 & 1.00 & 58811 &      \\
G44.644$-$0.516 & 19 14 53.76 & 10 10 07.69 & 2.55 & 58832 &      \\
G45.071+0.132 & 19 13 22.12 & 10 50 53.11 & 1.90 & 58832 &      \\
G45.380$-$0.594 & 19 16 34.14 & 10 47 01.60 & 0.85 & 58770 &      \\
G45.445+0.069 & 19 14 18.31 & 11 08 59.40 & 0.75 & 58706 &      \\
{\bf G45.467+0.053} & 19 14 24.15 & 11 09 43.00 & 0.85 & 58711 & 1, 4\\
G45.473+0.134 & 19 14 07.36 & 11 12 16.00 & 0.85 & 58725 &      \\
G45.493+0.126 & 19 14 11.35 & 11 13 06.20 & 1.50 & 58866 &      \\
G46.066+0.220 & 19 14 56.07 & 11 46 12.98 & 0.70 & 58814 &      \\
G46.115+0.387 & 19 14 25.52 & 11 53 25.99 & 0.70 & 58703 &      \\
G48.902$-$0.273 & 19 22 10.33 & 14 02 43.51 & 0.95 & 58719 &      \\
{\bf G48.990$-$0.299} & 19 22 26.13 & 14 06 39.78 & 0.80 & 58767 & 4\\
G49.417+0.324 & 19 20 59.82 & 14 46 49.10 & 0.75 & 58726 &      \\
{\bf G49.470$-$0.371} & 19 23 37.90 & 14 29 59.30 & 0.95 & 58810 & 3, 4\\
G49.617$-$0.360 & 19 23 52.81 & 14 38 03.30 & 1.65 & 58832 &      \\
G50.315+0.676 & 19 21 27.47 & 15 44 18.60 & 0.75 & 58802 &      \\
{\bf G50.779+0.152} & 19 24 17.41 & 15 54 01.60 & 1.75 & 58862 & 4\\
G51.679+0.719 & 19 23 58.87 & 16 57 41.80 & 1.50 & 58833 &      \\
G51.818+1.250 & 19 22 17.95 & 17 20 06.50 & 0.75 & 58770 &      \\
G52.922+0.414 & 19 27 34.96 & 17 54 38.14 & 1.70 & 58759 &      \\
G53.036+0.113 & 19 28 55.49 & 17 52 03.11 & 1.05 & 58731 &      \\
G53.142+0.071 & 19 29 17.58 & 17 56 23.21 & 0.70 & 58713 &      \\
G53.618+0.036 & 19 30 23.01 & 18 20 26.68 & 0.70 & 58704 &      \\
G56.963$-$0.235 & 19 38 17.10 & 21 08 05.40 & 0.75 & 58707 &      \\
G57.610+0.025 & 19 38 40.74 & 21 49 32.70 & 0.75 & 58719 &      \\
G58.775+0.644 & 19 38 49.13 & 23 08 40.20 & 0.80 & 58720 &      \\
G59.634$-$0.192 & 19 43 50.00 & 23 28 38.80 & 0.70 & 58712 &      \\
G59.833+0.672 & 19 40 59.33 & 24 04 46.50 & 0.85 & 58714 &      \\
G60.575+0.186 & 19 45 52.48 & 24 17 42.99 & 0.70 & 58706 &      \\
G69.539+0.975 & 20 10 09.07 & 31 31 34.86 & 0.70 & 58704 &      \\
G70.181+1.741 & 20 00 54.16 & 33 31 30.88 & 0.70 & 58705 &      \\
G71.522+0.385 & 20 12 57.91 & 33 30 26.95 & 0.75 & 58725 &      \\
G73.06+1.80 & 20 08 10.20 & 35 59 23.70 & 0.65 & 58705 &      \\
G75.782+0.342 & 20 21 44.05 & 37 26 36.91 & 0.70 & 58706 &      \\
G78.122+3.633 & 20 14 26.04 & 41 13 33.39 & 0.70 & 58705 &      \\
G78.886+0.708 & 20 29 24.94 & 40 11 19.28 & 1.40 & 58866 &      \\
G80.861+0.383 & 20 37 00.96 & 41 34 55.70 & 0.80 & 58705 &      \\
G81.722+0.571 & 20 39 01.05 & 42 22 49.18 & 0.80 & 58726 &  \\
G81.744+0.590 & 20 39 00.38 & 42 24 36.91 & 0.90 & 58707 &    \\
G81.752+0.590 & 20 39 01.99 & 42 24 59.08 & 0.80 & 58725 &      \\
G81.871+0.780 & 20 38 36.42 & 42 37 34.56 & 0.65 & 58704 &      \\
G90.921+1.486 & 21 09 12.98 & 50 01 03.56 & 0.75 & 58703 &      \\
{\bf G94.602$-$1.796} & 21 39 58.26 & 50 14 20.96 & 0.60 & 58703 &  2\\
G97.52+3.17 & 21 32 13.00 & 55 52 56.00 & 0.85 & 58707 &      \\
G98.035+1.446 & 21 43 01.43 & 54 56 17.75 & 0.75 & 58712 &      \\
G108.184+5.519 & 22 28 51.40 & 64 13 41.31 & 0.65 & 58706 &      \\
G108.766$-$0.986 & 22 58 51.18 & 58 45 14.37 & 0.95 & 58702 &      \\
G111.255$-$0.769 & 23 16 10.33 & 59 55 28.43 & 0.80 & 58702 &      \\
G121.298+0.659 & 00 36 47.35 & 63 29 02.16 & 0.75 & 58712 &      \\
G123.066$-$6.309 & 00 52 24.19 & 56 33 43.17 & 0.70 & 58712 &      \\
G136.845+1.167 & 02 49 33.59 & 60 48 27.95 & 0.85 & 58705 &      \\
G173.482+2.446 & 05 39 13.05 & 35 45 51.29 & 0.75 & 58705 &      \\
G173.70+2.89 & 05 41 37.40 & 35 48 49.00 & 0.70 & 58726 &     \\
G174.201$-$0.071 & 05 30 48.01 & 33 47 54.61 & 0.85 & 58726 &      \\
G188.793+1.030 & 06 09 06.96 & 21 50 41.23 & 0.80 & 58703 &      \\
G189.030+0.784 & 06 08 40.67 & 21 31 06.90 & 0.85 & 58705 &      \\
G189.471$-$1.216 & 06 02 08.37 & 20 09 20.10 & 0.80 & 58727 &      \\
G189.777+0.344 & 06 08 35.30 & 20 39 06.59 & 0.90 & 58727 &      \\
{\bf G196.454$-$1.677} & 06 14 37.05 & 13 49 36.16 & 0.80 & 58719 & 1 \\
\hline
\end{tabular}
\tablebib{
(1) \citet{caswell1995b}; (2) \citet{blaszkiewicz2004} (3) \citet{breen2010}; (4) \citet{breen2016}.
}
\end{table*}

\begin{table*}
\caption{Table with data used to create Fig. \ref{fig:breenvstmscpeak}. Column 2 presents crude estimation of the 6.7\,GHz maser peak flux density from \citet{breen2015}, column 3 presents peak flux densities from unpublished 6.7\,GHz line survey. Data for column 4 and 6 is taken from \citet{breen2016}, Table 2. Columns 5 and 7 present the same data, as in Table \ref{tab:dec_list}. Last two columns present number of visible features on 12\,GHz spectra from \citet{breen2016} and this survey, respectively. \label{tab:peak_comptab}}
\centering
\begin{tabular}[c]{c|cc|cccccc}
\hline
Source (l   b) & \multicolumn{2}{|c}{6.7\,GHz} & \multicolumn{5}{|c}{12.2\,GHz} \\
  & S$_{\mathrm{peak\,2010}}$ & S$_{\mathrm{peak\,2019}}$ & S$_{\mathrm{peak\,2010}}$ & S$_{\mathrm{peak\,2019}}$ & V$_{\mathrm{peak\,2010}}$ & V$_{\mathrm{peak\,2019}}$ & N$_{\mathrm{feat.\,2010}}$ & N$_{\mathrm{feat.\,2019}}$ \\
 (\degr \hspace{0.5cm} \degr) & (Jy) & (Jy) & (Jy) & (Jy) & (\kms) & (\kms) & & \\
\hline
G32.744$-$0.075 & 56.0  & 38.3  & 7.2  & 3.7  & 30.5  & 30.6  & 3 & 3 \\
G33.641$-$0.228 & 140.0 & 145.0 & 29.3 & 37.8 & 58.8  & 60.3  & 3 & 2 \\
G35.200$-$1.736 & 500.0 & 37.0  & 46.0 & 11.9 & 44.6  & 45.2  & 8 & 1 \\
G36.115+0.552   & 40.0  & 13.7  & 4.9  & 2.4  & 74.6  & 75.1  & 1 & 1 \\
G37.430+1.518   & 400.0 & 293.0 & 90.0 & 73.8 & 41.2  & 41.3  & 1 & 1 \\
G37.546$-$0.112 & 4.0   & 4.0   & 1.4  & 1.1  & 50.0  & 50.0  & 1 & 1 \\
G40.282$-$0.219 & 22.0  & 25.5  & 3.7  & 1.7  & 74.3  & 74.4  & 3 & 3 \\
G40.425+0.700   & 22.0  & 24.7  & 8.2  & 7.9  & 6.7   & 6.6   & 2 & 2 \\
G42.034+0.190   & 22.0  & 14.1  & 2.6  & 4.6  & 11.5  & 11.5  & 4 & 4 \\
G43.890$-$0.784 & 9.0   & 9.1   & 4.2  & 2.4  & 47.9  & 51.7  & 2 & 2 \\
G45.804$-$0.356 & 12.0  & 12.3  & 2.8  & 1.3  & 60.0  & 60.0  & 1 & 1 \\
G49.043$-$1.079 & 33.0  & 12.1  & 7.1  & 0.9  & 37.4  & 36.3  & 2 & 1 \\
G49.265+0.311   & 7.0   & 6.5   & 1.2  & 1.4  & -4.2  & -4.6  & 1 & 2 \\
G49.349+0.413   & 8.0   & 7.3   & 2.8  & 2.0  & 68.3  & 68.1  & 2 & 2 \\
G49.416+0.324   & 8.0   & 9.6   & 0.9  & 0.8  & -10.5 & -10.2 & 1 & 1 \\
G52.199+0.723   & 6.0   & 9.8   & 5.4  & 2.8  & 3.7   & 3.3   & 2 & 2 \\
G52.663$-$1.092 & 4.5   & 5.9   & 2.2  & 2.6  & 65.2  & 65.2  & 1 & 1 \\
\hline
\end{tabular}
\end{table*}

\clearpage

\begin{figure*}
\centering
\includegraphics[width=0.9\textwidth]{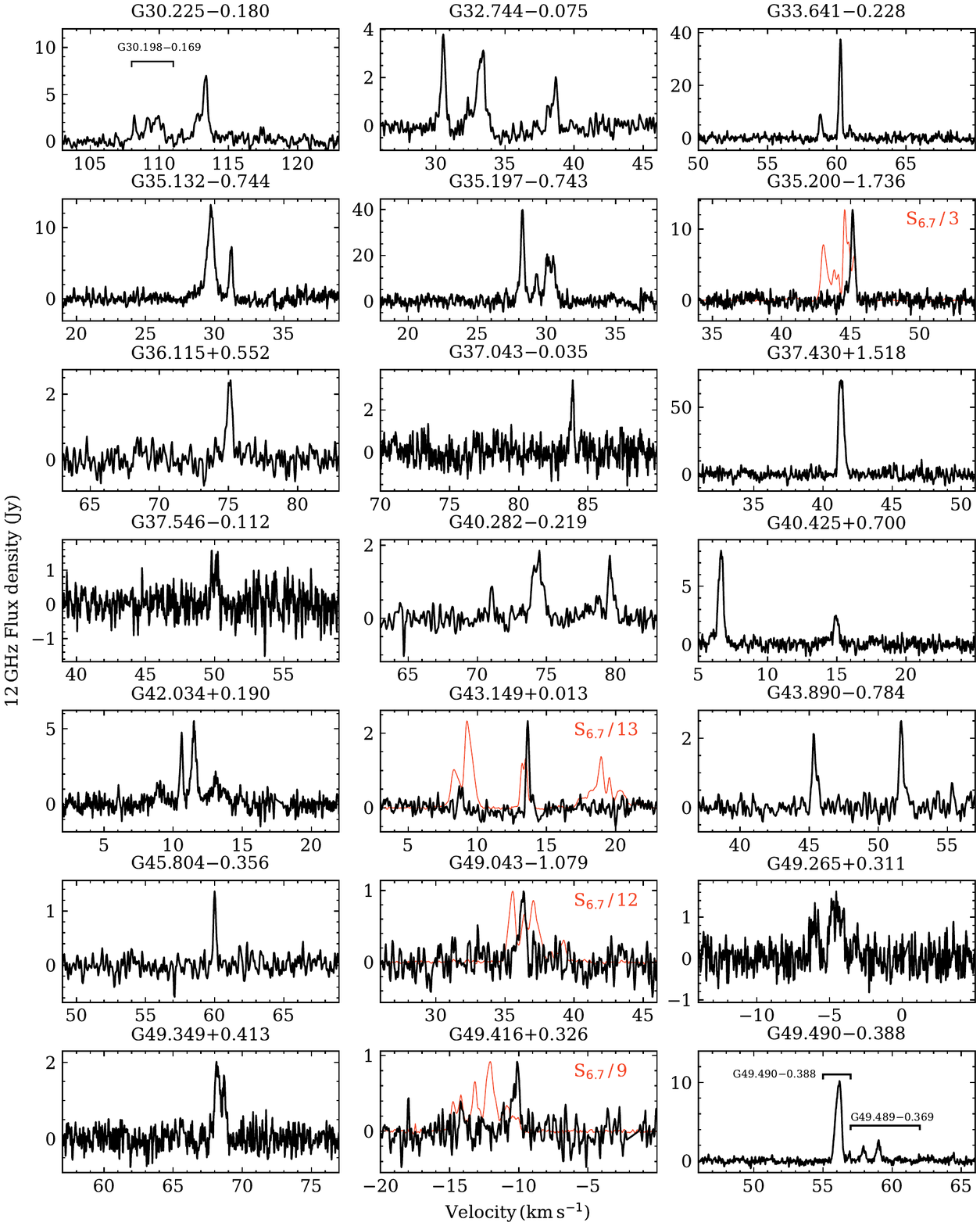}
\caption{Spectra of 12.2\,GHz (black) and 6.7\,GHz (red) methanol maser lines for previously known sources. The 6.7\,GHz spectrum is shown only if taken contemporaneously with that at 12.2\,GHz.
The flux density of 6.7\,GHz line is scaled by the factor given in the upper right corner. \label{fig:specrtas_nonewdec}}
\end{figure*}

\addtocounter{figure}{-1}
\begin{figure*}
\centering
\includegraphics[width=0.9\textwidth]{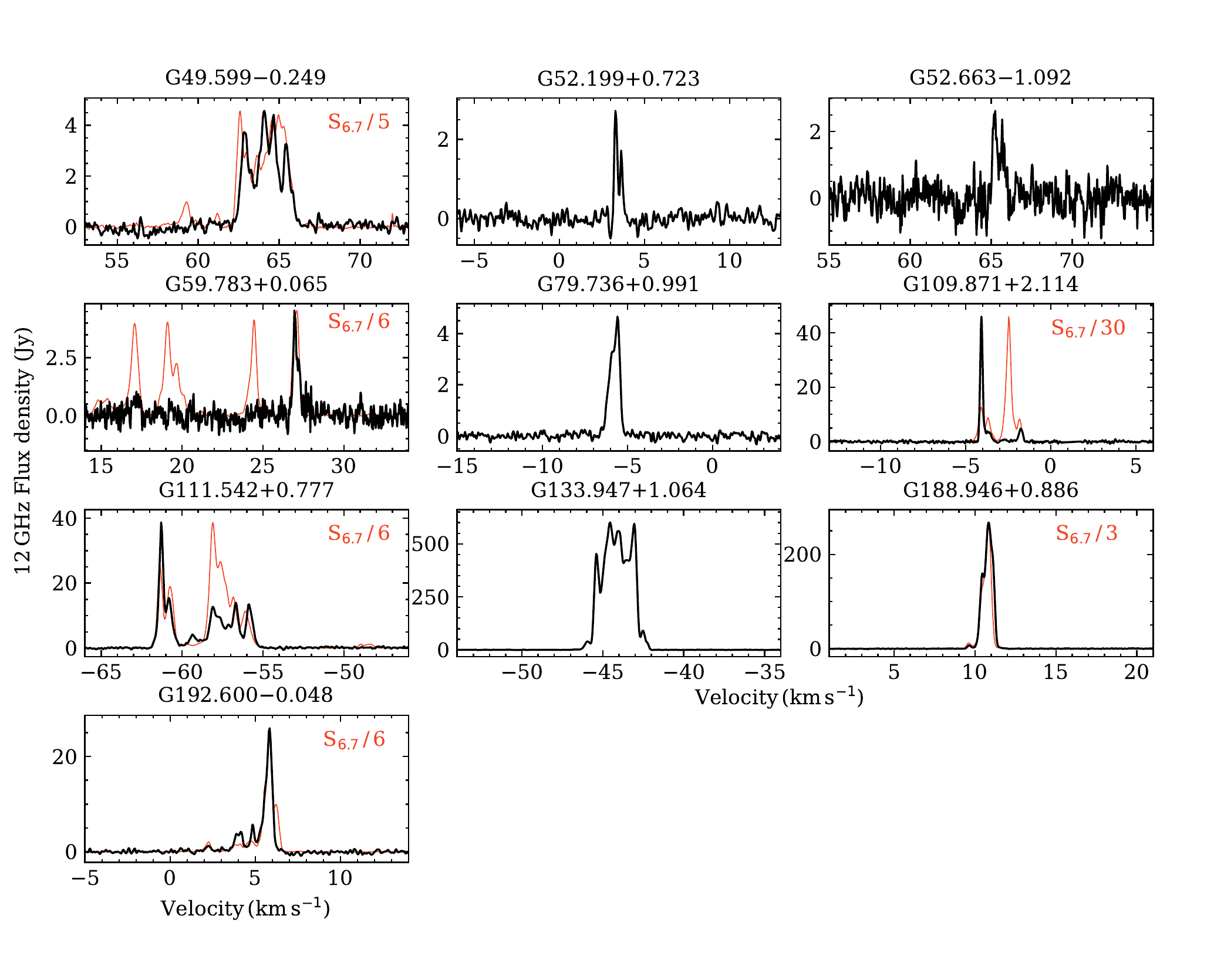}
\caption{continued}
\end{figure*}

\end{document}